\documentclass[preprintnumbers,amsmath,amssymb,eqsecnum]{revtex4}

\usepackage{epsf}
\usepackage{graphicx}  
\usepackage{dcolumn}   
\usepackage{bm}        

\begin{document}

\preprint{hep-th/0406040}

\title{ Non-asymptotically AdS/dS Solutions  and Their Higher
 Dimensional Origins}

 \author{Rong-Gen Cai\footnote{e-mail address:
cairg@itp.ac.cn; Rong-Gen$\_$Cai@baylor.edu}}

\address{CASPER, Department of Physics, Baylor University, Waco,
 TX76798-7316, USA \\
  Institute of Theoretical Physics, Chinese
Academy of Sciences,
 P.O. Box 2735, Beijing 100080, China}

\author{Anzhong Wang\footnote{e-mail:
Anzhong\_Wang@baylor.edu}}

\address{CASPER, Department of Physics, Baylor University, Waco,
 TX76798-7316, USA}

\begin{abstract}
We look for and analyze in some details some exact solutions of
Einstein-Maxwell-dilaton gravity with one or two Liouville-type
dilaton potential(s) in an arbitrary dimension. Such a theory
could be obtained by dimensionally reducing Einstein-Maxwell
theory with a cosmological constant to a lower dimension. These
(neutral/magnetic/electric charged) solutions can have a (two)
black hole horizon(s), cosmological horizon, or a naked
singularity. Black hole horizon or cosmological horizon of these
solutions can be a hypersurface of positive, zero or negative
constant curvature. These exact solutions are neither
asymptotically flat, nor asymptotically AdS/dS. But some of them
can be uplifted to a higher dimension, and those higher
dimensional solutions are either asymptotically flat, or
asymptotically AdS/dS with/without a compact constant curvature
space. This observation is useful to better understand holographic
properties of these non-asymptotically AdS/dS solutions.

\end{abstract}
\maketitle


\section{Introduction}
Looking for exact solutions of Einstein's field equations
with/without matter source is a subject of long standing interest.
Recent years have seen a lot of activity to find and study
asymptotically anti de Sitter (AdS) or de Sitter (dS) solutions.
For the asymptotically AdS spaces, there are at least two reasons
 responsible for this.  The first is related to the so-called
  AdS-CFT (conformal field theory) correspondence~\cite{AdS},  which
  states that  string/M theory on
 an AdS space times a compact manifold is dual to a strong coupling
 conformal field theory residing on the boundary of the AdS
 space. The radial spatial coordinate in asymptotically AdS spaces can be viewed as
 the energy scale in renormalization group flow of dual field theory.
 It has been argued by Witten~\cite{Witten} that the
 thermodynamics of black hole in AdS space can be identified with
 that of a certain dual CFT in the high temperature limit. For
 instance, the Hawking-Page phase transition~\cite{HP} of black holes in AdS
 space is identified with the confinement/unconfinement phase
 transition of CFT's.
 Therefore, one can gain some insights into the thermodynamic
 properties and phase structures of some strong coupling CFT's by
 investigating the thermodynamics of AdS black holes.

 The other is related to the topology structure of black hole in AdS space.
 Before the discovery of the AdS-CFT correspondence, it was already recognized that in
 an asymptotically AdS space, except for the black hole with event horizon being a positive
  constant curvature sphere, it is also possible to have black holes whose event horizon
  could be a negative constant or zero curvature hypersurface.
  These black hole solutions are often referred to as topological
  black holes in the literature~\cite{topo}. Various properties associated with
  these topological black holes have been investigated in recent years.
  For example, their higher dimensional~\cite{Birm} and charged \cite{charged}
   generalizations and  formation via gravitational collapse~\cite{coll} have
  been studied.  Topological black hole solutions in Lovelock
  gravity and associated thermodynamics have also been discussed~\cite{Lovelock}.
  Even some perturbative solutions in higher derivative gravity have been
  found and investigated in \cite{Nojiri}. In particular, topological black holes in gauged
  supergravities with nontrivial scalar fields have been also
  discovered in \cite{gauge}.

  In a $D$-dimensional Hilbert-Einstein action with a negative cosmological constant,
  $\Lambda = -(D-1)(D-2)/2l^2$,
  \begin{equation}
  \label{1eq1}
  S=\frac{1}{16\pi G_D}\int d^{D}x\sqrt{-g}\left(R +\frac{(D-1)(D-2)}{l^2}\right),
  \end{equation}
  where $G_D$ is the Newton gravitational constant in $D$ dimensions,
   one has a neutral AdS black hole solution with the metric
  \begin{equation}
  \label{1eq2}
  ds^2 = -f(r)dt^2 +f(r)^{-1} dr^2 + r^2 d\Sigma_{D-2,k}^2,
  \end{equation}
  where
  \begin{equation}
  \label{1eq3}
   f(r) =  k -\frac{16\pi G_D M}{(D-2) Vol(\Sigma)
      r^{D-3}}+\frac{r^2}{l^2},
      \end{equation}
 $M$ is  the mass of the black hole, and $d\Sigma^2_{D-2,k}$ represents the line
  element of a $(D-2)$-dimensional  hypersurface with
  constant curvature $(D-2)(D-3)k$ and volume $Vol(\Sigma)$. Without loss of generality, the
  characteristic curvature constant $k$ can be taken $\pm 1$ or $0$. When $k=1$, the
  hypersurface $\Sigma$ could be a $(D-2)$-dimensional round sphere
  $S^{D-2}$, while $k=0$, $\Sigma$ could be a $(D-2)$-dimensional torus
  $T^{D-2}$. When $k=-1$, the hypersurface $\Sigma$ is a negative
  constant curvature space. In this case, one can obtain a closed
  hyperbolic hypersurface $H^{D-2}$ with high genus through appropriate identification.
  In the AdS/CFT correspondence, the thermodynamics of black hole
  (\ref{1eq1}) can be viewed as that of the dual CFT residing
  on the boundary of the AdS space, whose metric is, up to a conformal factor,
  $ ds^2_{\rm CFT}=-dt^2 +l^2 d\Sigma_{D-2,k}^2. $
  In a word, the event horizon of the black hole (\ref{1eq1}) can
  have the topology $S^{D-2}$, $T^{D-2}$, or $H^{D-2}$, respectively, and the dual
  CFT resides on a $(D-1)$-dimensional spacetime with topology
   $R\times S^{D-2}$, $R\times T^{D-2}$, or $R\times H^{D-2}$, respectively.

For the dS space, similar to the AdS-CFT correspondence,
Strominger \cite{Stro} has argued that there is also  a dS-CFT
correspondence: quantum gravity on a dS space can be dual to a
Euclidean CFT residing on the boundary of the dS space. The time
coordinate in asymptotically dS spaces can be regarded as the
energy scale of renormalization group flow for dual Euclidean
field theory~\cite{Stro2,Bala}.  Replacing $ l^2$ by $-l^2$ and
$M$ by $-M$ in (\ref{1eq3}), the resulting solution is named as
the topological dS solution in~\cite{CMZ}. In that case, black
hole horizon disappears, instead a cosmological horizon occurs
with a cosmological singularity at $r=0$. The implication of such
asymptotically dS solutions  to the so-called mass bound
conjecture in dS space~\cite{Bala} has been investigated in
\cite{CMZ}.

   On the one hand, it is certainly of interest to find new, nontrivial solutions to
   the equations of motion of the action (\ref{1eq1}). On the other hand, let us note that
     various actions of gauged supergravities under consistent truncation,
    include  scalar fields with potentials of exponential form (Liouville-type).
    Sometimes the action also includes Maxwell fields. In a
    simpler form, the action could be written down as
    \begin{equation}
    \label{1eq4}
    S=\frac{1}{16\pi G_D} \int d^D x\sqrt{-g}\left( R- \frac{1}{2}
     (\partial \phi)^2 -2 \Lambda_0 e^{a\phi} -\frac{1}{4}e^{-b
     \phi}F_2^2\right),
     \end{equation}
     where $a$, $b$, and $\Lambda_0$ are constants and $F_2$
     denotes the Maxwell field. Usually people are interested in
     the asymptotically flat/AdS/dS solutions. However, it turns out
     difficult (might be impossible) to find such solutions in (\ref{1eq4}) with
     a set of general parameters. Indeed,
     in the literature, for example, see Refs.~\cite{Mann}-\cite{CZ}, some exact
     solutions of the action (\ref{1eq4}) have been found in some special cases.
    These solution are asymptotically neither AdS nor dS,
     different from
     those (\ref{1eq2}) which are asymptotically AdS or dS. We have studied
     some aspects ~\cite{CO,CZ,CMZ} of dual field theory to these gravity solutions
     in the spirit of holography, as the non-conformal
     generalizations of AdS-CFT correspondence and dS-CFT
     correspondence.

     Note that an action like (\ref{1eq4}) can also be obtained by dimensionally reducing
      Einstein-Maxwell theory with a cosmological constant to a lower
     dimension.  It is certainly worth discussing the relationship between those
     non-asymptotically AdS/dS solutions of the action (\ref{1eq4})
      and asymptotically AdS/dS solutions (\ref{1eq2}), in order to better understand
     the holographic properties of those non-asymptotically AdS/dS
     solutions. This is one of the main aims of the present
     paper~\footnote{While this paper was writing up, an
     interesting paper \cite{CL} appears (see also
     \cite{CGL}). The
     authors find that a four dimensional non-asymptotically
     flat black hole solution derived by Chan et al.~\cite{Mann}
     of Einstein-Maxwell-dilaton theory (\ref{1eq4}) without the
     dilaton potential term can be reobtained by taking a near
     horizon limit of asymptotically flat black hole solution, and
     that when $b=\sqrt{6}$ (in the notation of \cite{CL},
     $\alpha=\sqrt{3}$), the four dimensional non-asymptotically
     flat, magnetic charged black hole can be understood as a
     dimensional reduction of a five dimensional, asymptotically flat
     Schwarzschild black hole along one of the azimuthal angles.}.
     On the other hand, we will look for and analyze in
     some details, exact solutions in an action like
     (\ref{1eq4}) with one or two Liouville-type potentials,
     without any assumption on the relations among the parameters
     in the theory, which generalizes existing solutions in the literature
     in various directions.

     The organization of the present paper is as follows. In the
     next section, we  first discuss the case where resulting action
     comes from a higher dimensional Einstein-Maxwell theory with a
     cosmological constant and the reduced
     subspace is Ricci flat. In this case, the resulting action is
     of the form (\ref{1eq4}). In Sec.~III, we investigate the
     case where the reduced subspace is a nonzero
     constant curvature space, where resulting action has two
     Liouville-type potentials. The conclusions and some discussions are
     presented in Sec.~IV.

\section{Dimensional reduction: Ricci flat case}

\subsection{Neutral solution}
 In this subsection, we first discuss the case without the Maxwell field.
 Let us start with a $(D+d)$-dimensional Hilbert-Einstein action
 with a cosmological constant $\Lambda_0$,
 \begin{equation}
 \label{2eq1}
 S =\frac{1}{16\pi G}\int d^{D+d}x\sqrt{- \tilde g}(\tilde R -2
 \Lambda_0),
 \end{equation}
 where $G$ is the Newton gravitational constant in $D+d$ dimensions,
 $\tilde g$ and $\tilde R$ denote the $(D+d)$-dimensional metric
 determinant and curvature scalar, respectively. Consider the
 $(D+d)$-dimensional metric line element having the following form
 \begin{equation}
 \label{2eq2}
 ds^2_{D+d}= e^{2\alpha (x)} ds_D^2 + e^{2\beta(x)}ds_d^2,
 \end{equation}
 where $ds^2_d=q_{ij}dy^idy^j$ denotes a $d$-dimensional
 constant curvature space with scalar curvature $d(d-1)k_d$.
 In this section, we consider the case $k_d=0$.
  Namely, $q_{ij}=\delta_{ij}$, and then $ds_d^2$ is a $d$-dimensional Euclidean
  flat space. In addition, $\alpha$ and $\beta$ are two
 functions of coordinate $x$ in the subspace described by
 $ds_D^2$. Making dimensional reduction along $y$, and taking
 $\beta=(2-D)\alpha/d$, we obtain an  action from
 (\ref{2eq1})
 \begin{equation}
 \label{2eq3}
 S=\frac{V_q}{16\pi G}\int d^Dx\sqrt{-g}\left ( R-
 \frac{(D-2)(D+d-2)}{d}(\partial \alpha)^2 -2 \Lambda_0
 e^{2\alpha}\right),
 \end{equation}
 where $V_q$ is the volume of the subspace described
 by $ds^2_d$ and $R$ is the curvature scalar of
 line element $ds_D^2$. Thus $G/V_q$ stands for the effective Newton gravitational
 constant in $D$ dimensions here.  Defining
 \begin{equation}
 \label{2eq4}
 \alpha = \sqrt{\frac{d}{2(D-2)(D+d-2)}} \phi,
 \end{equation}
 the kinetic term of the scalar field is changed to
 have a canonical form
 \begin{equation}
 \label{2eq5}
S=\frac{V_q}{16\pi G}\int d^Dx\sqrt{-g}\left ( R-
 \frac{1}{2}(\partial \phi)^2 -2 \Lambda_0
 e^{a \phi}\right),
 \end{equation}
 where
\begin{equation}
\label{2eq6}
  a = \sqrt{\frac{2d}{(D-2)(D+d-2)}}.
  \end{equation}
 The action (\ref{2eq5}) is just the one (\ref{1eq4}) with
 $F_2=0$. Varying the action (\ref{2eq5}) yields equations
 of motion
 \begin{eqnarray}
 \label{2eq7}
 && R_{\mu\nu}=\frac{1}{2}\partial_{\mu}\phi \partial _{\nu}\phi
 +\frac{2}{D-2}\Lambda_0 e^{a\phi}g_{\mu\nu}, \nonumber \\
 && \nabla^2\phi -2a \Lambda_0 e^{a\phi}=0.
 \end{eqnarray}
 Now we solve the equations of motion (\ref{2eq7}). In the process of solving
 these equations, for the sake of generality, we only assume that $a$ is
 a positive constant (this assumption always holds because it can be made
 positive via $\phi \to -\phi$ if it was negative) without
 imposing the condition (\ref{2eq6}). Suppose that the metric is of the
 form,
 \begin{equation}
 \label{2eq8}
 ds^2_D= A(r) dt^2 +A(r)^{-1}dr^2 +R(r)^2h_{mn}dx^mdx^n,
 \end{equation}
 where $h_{mn}dx^mdx^n$ denotes the line element of
 a $(D-2)$-dimensional hypersurface with constant curvature
 $(D-2)(D-3)k$ and volume $V_h$. Corresponding to the metric (\ref{2eq8}),
 we can write the equations of motion as
 \begin{eqnarray}
 \label{2eq9}
 && R^t_t = -\frac{A''}{2} -(D-2) \frac{A'R'}{2R} =
 \frac{2}{D-2}\Lambda_0 e^{a\phi}, \\
 \label{2eq10}
 && R^r_r = -\frac{A''}{2} -(D-2)\frac{A'R'}{2R}
 -(D-2)\frac{AR''}{R}=\frac{1}{2}A\phi'^2
 +\frac{2}{D-2}\Lambda_0 e^{a\phi}, \\
 \label{2eq11}
 && R^m_n = \delta^m_n\left\{\frac{D-3}{R^2}k
 -\frac{1}{(D-2)R^{D-2}}[A(R^{D-2})']'\right\} =\frac{2}{D-2}\Lambda_0
 e^{a\phi}\delta^m_n, \\
 \label{2eq12}
 && \frac{1}{R^{D-2}}\left(R^{D-2}A
 \phi'\right)'-2a\Lambda_0e^{a\phi}=0,
 \end{eqnarray}
where a prime stands for the derivative with respect to $r$. From
(\ref{2eq9}) and (\ref{2eq10}), one has
\begin{equation}
(D-2)\frac{R''}{R}= -\frac{1}{2}\phi'^2.
\end{equation}
Setting $R=r^N$, where $N$ is a constant to be determined shortly,
one then has
\begin{equation}
\label{2eq14} \phi'^2 = -2N(N-1)(D-2)/r^2.
\end{equation}
Integrating this and substituting into
Eqs.(\ref{2eq9})-(\ref{2eq12}), we find solutions to the equations
(\ref{2eq9})-(\ref{2eq12}), depending on the characteristic
curvature constant $k$.

(1). When $k=0$, we have
\begin{eqnarray}
\label{2eq15}
 && A(r)= -\frac{{\cal M}}{r^{(D-2)N-1}} -\frac{2 \Lambda_0
 e^{a\phi_0}r^{2N}}{N(ND-1)(D-2)},
   \\
 && \phi =\phi_0 -\sqrt{2N(1-N)(D-2)}\ln r, \ \ \
 R =r^N, \nonumber \\
 \label{2eq16}
 && N =\frac{2}{2+a^2(D-2)},
 \end{eqnarray}
 where ${\cal M}$ and $\phi_0$ are two integration constants. The solution
 (\ref{2eq15}) has a singularity at $r=0$ only.  From
 (\ref{2eq14}) and (\ref{2eq16}), one can see that $0<N<1$. When
 $a=0$, one has $N=1$ and the $\phi$ is then a constant.  In this
 case, the action (\ref{2eq5}) reduces to (\ref{1eq1}) and the solution
 (\ref{2eq15}) becomes (\ref{1eq2}) with $k=0$.

 We are of course interested in the case of $a\ne 0$ and $\Lambda_0 <0$.
 In this case, the dilaton potential in (\ref{2eq5}) could be
 viewed as a negative effective cosmological constant.
 (i) when $1/D <N <1$, the second term in (\ref{2eq15}) is positive and
 dominant as $r \to \infty$. The solution therefore describes a black
 hole, provided ${\cal M}>0$,
 with a conformal Ricci flat horizon $r_+$ determined by $A(r)|_{r=r_+}=0$.
 Applying the quasilocal mass formulism of gravitational
 configurations developed in \cite{quasi} to our case and considering
 the solution with ${\cal M}=0$ as the reference background,
 the quasilocal mass of the black hole is found to be (see also
 \cite{CO})
 \begin{equation}
 \label{2eq17}
 M =\frac{(D-2) N V_q V_h {\cal M}}{16\pi G }.
 \end{equation}
 Note that here the effective Newton gravitational constant is $G/V_q$.
 Obviously, if ${\cal M}<0$, the singularity at $r=0$ becomes
 naked.
 (ii) When $0<N<1/D$, the  second term is negative and the first
 term in (\ref{2eq15}) is dominant as $r\to \infty$. In this
 case, the solution has a black hole horizon $r_+$ provided ${\cal M}<0$ (which
 implies that the gravitational mass of the black hole is negative).
  The horizon is also determined  by
 $A(r)|_{r=r_+}=0$  and is a conformal Ricci flat hypersurface. When ${\cal
 M}>0$, the  singularity at $r=0$ becomes naked, again.

 For completeness, we also analyze the case of $\Lambda_0>0$ here. In
 this case, the dilaton potential in (\ref{2eq5}) can be regarded
 as a positive effective cosmological constant.
 (i) When $1/D <N<1$, the second term in (\ref{2eq15}) is always negative and
 dominant as $r\to \infty$, therefore there is a cosmological horizon provided
 ${\cal M}<0$. The solution describes a cosmological solution with a conformal Ricci flat
 cosmological horizon $r_c$ determined by $A(r)|_{r=r_c}=0$, and the singularity at $r=0$
 is a cosmological singularity. The gravitational mass of
the solution is found to be (see also \cite{CMZ})
 \begin{equation}
 \label{2eq18}
M =-\frac{(D-2)N V_q V_h {\cal M}}{16\pi G },
 \end{equation}
 in the sense of \cite{Bala}. When ${\cal M}>0$, the solution
 describes a naked singularity spacetime. This solution was first
 found in \cite{CMZ} and the implication of the
 solution in the sense of dS/CFT correspondence was
 discussed there. (ii) When $0<N <1/D$, the second term is always positive
 and the first term is dominant as $r\to \infty$.  The solution
 has a cosmological horizon if ${\cal M}>0$, otherwise
 describes a naked singularity.  The property of the
 solution is summarized in Tab.~I.

\begin{table}[ht]
\caption{The property of the solution (\ref{2eq15}) for different
parameters. Here ``BH" means that the outmost horizon of the
solution is a black hole horizon, ``CH"  a cosmological horizon,
and ``NK" stands for that the singularity at $r=0$ is naked. The
same notation will be used below.} \label{tab:1}
\begin{center}
\begin{tabular}{l|c|c|c|c|c|c|c|c}
\hline \hline
 $\Lambda_0 $ &$<0$ & $<0$  & $ <0 $ & $ <0 $ & $ >0$   & $ >0$ & $ >0$ & $>0$ \\
\hline
 $N$ & $0<N<1/D$ &$0<N<1/D$ & $1/D <N<1$  & $1/D<N<1$  & $1/D<N<1$ &
$1/D <N<1 $ &
    $ 0<N< 1/D $ & $ 0<N<1/D$  \\
 \hline
  ${\cal M} $ & $ <0$ &$>0$ & $ >0 $ & $ <0 $ &  $<0$  & $>0$ & $ >0$ & $<0$ \\
  \hline
  Solution &  BH & NK & BH  & NK &  CH  & NK &  CH & NK \\
\hline \hline
\end{tabular}
\end{center}
\end{table}

 Note that  regardless of the values of parameters ${\cal M}$,
 $\Lambda_0$ and $N$, the spacetime (\ref{2eq8}) with
 (\ref{2eq15}) and (\ref{2eq16}) is neither asymptotically dS,
 nor AdS.  Now let us observe an interesting consequence if the
 parameter $a$ in (\ref{2eq15}) has a relation (\ref{2eq6}). In
 this case, one has $N= (D+d-2)/(D+2d-2)$, which obeys $1/2<N<1$.
 We can uplift the $D$-dimensional, non-asymptotically AdS/dS solution
 (\ref{2eq15}) to the one (\ref{2eq2}) in $ (D+d)$ dimensions
 and obtain
 \begin{equation}
 \label{2eq19}
 ds^2_{D+d}=e^{a\phi_0} r^{-2d/(D+2d-2)}\left(-Adt^2 +A^{-1}dr^2\right)
 +r^{2(D-2)/(D+2d-2)}(e^{a\phi_0}\delta_{mn}dx^mdx^n + e^{-(D-2)a\phi_0/d} \delta_{ij}dy^idy^j).
 \end{equation}
 Redefining $r^{(D-2)/(D+2d-2)} \to r$ and rescaling
 coordinates and the integration constant ${\cal M}$, we find that the solution
  (\ref{2eq19}) can be rewritten as
 \begin{equation}
 \label{2eq20}
 ds^2_{D+d} = -f(r) dt^2
   +f(r)^{-1} dr^2
   +r^2 (\delta_{mn}dx^mdx^n +\delta_{ij}dy^idy^j),
   \end{equation}
   with
   $$ f(r) =-\frac{{\cal M}}{r^{D+d-3}} -
   \frac{2\Lambda_0r^2}{(D+d-1)(D+d-2)}.$$
   This is nothing but the AdS black hole solution (\ref{1eq2}) with $k=0$ in $(D+d)$
   dimensions if $\Lambda_0 <0$ and ${\cal M}>0$. On the other
   hand, if $\Lambda_0>0$ and ${\cal M}<0$, the solution is a
   topological dS solution with $k=0$ in (\ref{1eq3}). Clearly the solution
   (\ref{2eq20}) is asymptotically AdS/dS.  Therefore some of
   non-asymptotically AdS/dS solutions (\ref{2eq15}) can be
   understood as the dimensional reduction of a higher
   dimensional, asymptotically AdS/dS solution (\ref{2eq20}).
   Certainly this observation is useful to understand the
   holography of the non-asymptotically AdS/dS solutions (\ref{2eq15}).

 When $k\ne 0$, the solution of (\ref{2eq7}) is found to be
\begin{eqnarray}
\label{2eq21}
 && A(r)= -\frac{{\cal M}}{r^{(D-2)N-1}} + \frac{(D-3)k
 r^{2-2N}}{(2N-1)(N(D-4)+1)}, \\
 && \phi =\phi_0 -\sqrt{2N(1-N)(D-2)}\ln r, \ \ \ R=r^N, \nonumber \\
 && N= \frac{a^2 (D-2)}{2+a^2(D-2)}, \nonumber\\
 \label{2eq22}
 && 2\Lambda_0 = -\frac{(D-3)(D-2) (1-N)k e^{-a\phi_0}}{2N-1}.
 \end{eqnarray}
It might be worth noticing that this set of solution with $k\ne 0$
does not reduce to the one (\ref{2eq15}) and (\ref{2eq16}) when $k
\to 0$. This is caused that in the process of solving the
equations (\ref{2eq9})-(\ref{2eq12}), by use of  $R=r^{N}$ and
$\phi$ given by (\ref{2eq14}), we can obtain a general solution of
$A$ through (\ref{2eq12}). In order the equation (\ref{2eq11}) to
hold, we have to have two constraints if $k\ne 0$, which result in
$N$ and $\Lambda_0$ in (\ref{2eq22}). If $k=0$, there is only one
constraint, giving us $N$ in (\ref{2eq16}). Therefore within the
ansatz of $R=r^{N}$, the solutions of $k=0$ and $k\ne 0$ belong to
different branches. In addition,  note that due to $0<N<1$, the
second term in (\ref{2eq21}) therefore is always dominant as
 $r\to \infty$.

 (2). When $k=1$, one can see from (\ref{2eq22}) that the
 cosmological constant $\Lambda_0$ is negative as $ 1/2<N<1 $. In
 this case, the solution (\ref{2eq21}) describes a black hole with
 mass (\ref{2eq17}) provided ${\cal M}>0$, whose horizon is determined
 by $A(r)|_{r=r_+}=0$, and the horizon is a $(D-2)$-dimensional, positive constant
 curvature hypersurface. If ${\cal M}<0$, it is a naked singularity solution.
 When $0<N<1/2 $, the cosmological constant has to be positive. In
 that case, the solution has a cosmological horizon determined by
 $A(r)|_{r=r_c}=0$ provided ${\cal M}<0$. Otherwise, it is a naked singularity solution.
 The gravitational mass is  given by (\ref{2eq18}). This solution generalizes the
 topological dS solutions in \cite{CMZ}. The properties of the solution are
 summarized in Tab.~II, where the cases of naked singularity
are not included.

\begin{table}[ht]
\caption{The property of the solution (\ref{2eq21}) for different
parameters. }
 \label{tab:2}
\begin{center}
\begin{tabular}{l|c|c|c|c}
\hline \hline
 $k$ & $1$ & $1$ & $-1$ & $-1$ \\ \hline
 $N$ & $0<N<1/2$ & $ 1/2 <N<1$ & $ 0<N<1/2$ & $1/2 <N<1$ \\
 \hline
 $\Lambda_0 $ & $>0$  & $ <0 $ & $<0$ & $>0$ \\
  \hline
  Solution & CH (${\cal M}<0$) & BH (${\cal M}>0$)
  & BH (${\cal M}>0)$ & CH (${\cal M}<0)$ \\
\hline \hline
\end{tabular}
\end{center}
\end{table}

 Where $a$ takes the value of (\ref{2eq6}), we can uplift the
  solution (\ref{2eq21})-(\ref{2eq22}) to the one in
  $(D+d)$-dimensions. In that case, one has $N = d/(D+2d-2)<1/2$.
  Namely we have to take a positive cosmological constant
  $\Lambda_0$ in (\ref{2eq22}). Upon  lifting, we obtain
   \begin{equation}
   ds^2_{D+d}= e^{a\phi_0}r^{-2N}(-A(r)dt^2 +A^{-1}(r)dr^2)+
        e^{-(D-2)a\phi_0/d} r^{2N(D-2)/d}\delta_{ij}dy^idy^j
        + e^{a\phi_0}h_{mn}dx^mdx^n.
\end{equation}
Redefining $r^{N(D-2)/N} \to r$ and rescaling the coordinates and
the integration constant ${\cal M}$, the solution can be rewritten
as
\begin{equation}
\label{2eq24}
ds^2_{D+d}= - f(r) dt^2 +f(r)^{-1}dr^2 +r^2
\delta_{ij}dy^idy^j + {\cal B}^2h_{mn}dx^mdx^n,
\end{equation}
where
$$ f(r)= -\frac{{\cal M}}{r^{d-1}}- \frac{2\Lambda_0 r^2}{(d+1)(D+d-2)},
\ \ \ {\cal B}^2 =\frac{(D-3)(D+d-2)}{2\Lambda_0},
$$
and the scalar curvature of the $(D-2)$-dimensional positive
constant curvature space $h_{mn}dx^mdx^n$ has been normalized to
$(D-2)(D-3)$. If ${\cal M}<0$, this solution is nothing but a
$(d+2)$-dimensional topological dS solution with a conformal Ricci
flat cosmological horizon times a $(D-2)$-dimensional compact
space with a positive constant curvature
$2(D-2)\Lambda_0/(D+d-2)$. It is easy to check that the solution
({\ref{2eq24}) satisfies the $(D+d)$-dimensional Einstein's field
equations with a positive cosmological constant $\Lambda_0$.

(3). When $k=-1$, one can see from (\ref{2eq22}) that the
cosmological constant is positive if $1/2 <N <1$. In this case,
the second term is negative and  the solution (\ref{2eq21})
 has a cosmological horizon which is a negative constant curvature
 hypersurface provided ${\cal M}<0$. The gravitational mass is
 given by (\ref{2eq18}). The solution generalizes the one
 in \cite{CMZ} to the case $k=-1$.
If ${\cal M}>0$, the solution describes a naked singularity. On
the other hand, when $0<N<1/2$, the cosmological constant has to
be negative. In this case, the second term in (\ref{2eq21}) is
positive and the solution describes a black hole with a negative
constant curvature horizon provided ${\cal M}>0$. The black hole
mass is given by (\ref{2eq17}). If ${\cal M}<0$, the singularity
at $r=0$ becomes naked. The causal structure of the solution is
summarized in Tab.~II.

When $a$ takes the value of (\ref{2eq6}), we have  $N =
d/(D+2d-2)<1/2$. Namely, in this case, one has to have a negative
cosmological constant. One can uplift the solution to $(D+d)$
dimensions. As the case of $k=1$, we find that the
$(D+d)$-dimensional solution can be rewritten as
\begin{equation}
\label{2eq25}
  ds^2_{D+d}= -f(r)dt^2 +f(r)^{-1}dr^2 + r^2 \delta_{ij}dy^idy^j
         +{\cal B}^2 h_{mn}dx^mdx^n,
         \end{equation}
where
$$ f(r) = -\frac{{\cal M}}{r^{d-1}} +
\frac{2|\Lambda_0|r^2}{(d+1)(D+d-2)}, \ \ \ {\cal
B}^2=\frac{(D-3)(D+d-2)}{2|\Lambda_0|}. $$
 Note that here
$h_{mn}dx^mdx^n$ denotes a $(D-2)$-dimensional negative constant
curvature space with scalar curvature $-(D-2)(D-3)$. It is easy to
see that the solution (\ref{2eq25}) describes a
$(d+2)$-dimensional AdS black hole with a conformal Ricci flat
horizon times a $(D-2)$-dimensional negative constant curvature
space. As the case of (\ref{2eq24}), the solution (\ref{2eq25})
obeys the $(D+d)$-dimensional Einstein's field equations with a
negative cosmological constant $\Lambda_0$.

\subsection{Charged solution}
Next let us add a term of Maxwell field to the action
(\ref{2eq1}),
\begin{equation}
\label{2eq26}
S=\frac{1}{16\pi G} \int d^{D+d} x \sqrt{-\tilde g}
 \left (\tilde R -2\Lambda_0 -\frac{1}{4} F_2^2\right).
 \end{equation}
 We consider magnetic charged solutions of the action (\ref{2eq26}).
  Then as the case without
 the Maxwell field, one can make a dimensional reduction according
 to (\ref{2eq2}) and obtain
 \begin{equation}
 \label{2eq27}
 S=\frac{V_q}{16\pi G} \int d^Dx\sqrt{-g} \left( R -\frac{1}{2}
  (\partial \phi)^2 -2\Lambda_0 e^{a\phi} -\frac{1}{4} e^{-b\phi} F_2^2
   \right),
   \end{equation}
  which has the same form as (\ref{1eq4}).
   Here $\phi$ has the same relation to $\alpha$ as the case
   without the Maxwell field (\ref{2eq4}), and
   \begin{equation}
   \label{2eq28}
   b = a = \sqrt{\frac{2d}{(D-2)(D+d-2)}}.
\end{equation}
 Varying the action (\ref{2eq27})
yields the equations of motion
\begin{eqnarray}
\label{2eq29}
 && R_{\mu\nu}=\frac{1}{2}(\partial_{\mu}
 \phi)(\partial_{\nu}\phi)  + \frac{2}{D-2}\Lambda_0 e^{a\phi}
 g_{\mu\nu} +\frac{1}{2}e^{-b \phi}\left(F_{\mu\lambda}F_{\nu}^{\
 \lambda}- \frac{1}{2(D-2)}F^2_2 g_{\mu\nu}\right), \nonumber \\
 && \partial_{\mu}(\sqrt{-g} e^{-b\phi} F^{\mu\nu})=0, \ \ \
 F_{\mu\nu,\lambda}+F_{\nu\lambda,\mu} +F_{\lambda\mu,\nu}=0,
  \nonumber \\
  && \nabla^2\phi -2a \Lambda_0 e^{a\phi}
  +\frac{b}{4}e^{-b\phi}F^2=0.
  \end{eqnarray}
Now we find magnetic charged solution of action (\ref{2eq27}). In
the process of solving these equations, we just keep $a$ and $b$
as two constants, and do not assume that they have a relation of
(\ref{2eq28}). In addition, for completeness we will also find
electric charged solutions of action (\ref{2eq27}).

\subsubsection{Magnetic charged solution}
Here it should be mentioned that for magnetic charged solutions in
(\ref{2eq26}), the dimensional reduction from (\ref{2eq26}) to
(\ref{2eq27}) should go to $D=4$. With the ansatz (\ref{2eq8}), we
find  several sets of solutions, which depend on the
characteristic curvature $k$,

(1). When $k=0$, we find two  sets of solutions to (\ref{2eq29}).
\begin{itemize}
 \item Solution 1
 \begin{eqnarray}
\label{2eq30}
 && A(r) =-\frac{\cal M}{r^{2N-1}} -\frac{\Lambda_0 e^{a\phi_0}
          r^{2N}}{N (4N-1)} +\frac{{\cal Q}^2 e^{-b\phi_0}}{4N r^{2N}},
 \\
&& \phi =\phi_0 -\sqrt{4N(1-N)}\ln r,\ \ \  R=r^N, \nonumber \\
&& N= \frac{1}{1+a^2}, \ \ \ b=-a, \nonumber \\
&& F_{mn} = {\cal Q} \epsilon_{mn},
\end{eqnarray}
where $\epsilon_{mn}$ is the volume element of $h_{mn}dx^mdx^n$
and ${\cal Q}$ is an integration constant related to the magnetic
charge $Q$ via $Q= {\cal Q} V_h/4\pi$. (i) When $0<N<1/4$, the
first term in (\ref{2eq30}) is dominant as $r \to \infty$.
Therefore there is a cosmological horizon if ${\cal M}>0$
regardless of the sign of $\Lambda_0$ provided ${\cal Q}\ne 0$
(here and after when a magnetic/electric charge is present, ${\cal
Q}\ne0$ is always assumed), while a black hole appears if ${\cal
M} <0$ and $\Lambda_0 <0$. In other cases, the singularity at
$r=0$ is naked. (ii) when $1/4 <N<1$, the second term in
(\ref{2eq30}) is always dominant as $r \to \infty$. Black hole
horizon will appear if ${\cal M}>0$ and $\Lambda_0 <0$. In this
case, the solution may have two or one black hole horizon or a
naked singularity depending on the parameters ${\cal M}$, ${\cal
Q}$ and $\Lambda_0$.  For instance, if $N=1/2$, we have two black
hole horizons
$$ r_{\pm}= \frac{e^{-a\phi_0}}{2|\Lambda_0|} \left ({\cal M}
 \pm \sqrt{{\cal M}^2 -2|\Lambda_0| {\cal Q}^2 e^{2a\phi_0}}\right)$$
 provided ${\cal M}^2 - 2|\Lambda_0|{\cal Q}^2 e^{2a\phi_0} >0$.
  If $\Lambda_0 >0$, a cosmological horizon occurs again in spite of the sign of ${\cal M}$.
 The causal structure of the solution (\ref{2eq30}) is summarized in Tab.~III. Due to
 the relation
$b=-a$ for this solution, which does not satisfy $b=a$ of
(\ref{2eq28}). Therefore this solution cannot be uplifted to the
case of $D+d$ dimensions according to (\ref{2eq2}) unless ${\cal
Q}=0$ . In the case of ${\cal Q}=0$, it results in the solution
(\ref{2eq20}).

\begin{table}[ht]
\caption{The property of the solution (\ref{2eq30}) for different
parameters. }
 \label{tab:3}
\begin{center}
\begin{tabular}{l|c|c|c|c}
\hline \hline
 $N$ & $0<N<1/4$ & $0 <N< 1/4$& $ 1/4 <N<1$ & $1/4 <N<1$ \\
 \hline
 $\Lambda_0 $ & $-$  & $ <0 $ & $ <0 $ & $>0$ \\
  \hline
  Solution &  CH (${\cal M}>0$)
   & BH (${\cal M}<0$) &
   BH (${\cal M} >0$) & CH   \\
\hline \hline
\end{tabular}
\end{center}
\end{table}

\item Solution 2
\begin{eqnarray}
\label{in1}
 && A = -\frac{{\cal M}}{r^{2N-1}} +\frac{{\cal Q}^2
e^{-b\phi_0}}{2(1-2N)}r^{2-2N}, \\
&& \phi =\phi_0 -\sqrt{4N(1-N)}\ln r, \ \ \ R=r^N, \nonumber \\
&& N = \frac{a^2}{1+a^2}, \ \ \  b=a, \nonumber \\
&& {\cal Q}^2 e^{-b\phi_0}=4(2N-1)\Lambda_0 e^{a\phi_0},
\end{eqnarray}
The second term in (\ref{in1}) is always dominant as $r\to
\infty$. The solution describes a black hole when $0<N<1/2$ and
${\cal M}>0$. When $1/2 <N<1$ and ${\cal M}<0$, the solution has a
cosmological horizon. In other cases, the singularity at $r=0$ is
naked. This solution is a special case of the solution 3 given by
(\ref{2eq36})-(\ref{2eq37}) below.

\end{itemize}

(2). When $k\ne 0$, we find three sets of solutions to the
equations of motion (\ref{2eq29}).
\begin{itemize}
\item Solution 1.
\begin{eqnarray}
\label{2eq32}
 && A =-\frac{\cal M}{r^{2N-1}} +\frac{k}{2N-1}r^{2-2N}
         +\frac{{\cal Q}^2 e^{-b \phi_0}}{4N
          r^{2N}}, \\
 && \phi =\phi_0 -\sqrt{4N(1-N)}\ln r, \ \ \  R =r^N, \nonumber \\
 && N  = \frac{a^2}{1+a^2}, \ \ \  b=-\frac{1}{a}, \nonumber \\
 && \Lambda_0 = -\frac{(1-N)k e^{-a\phi_0}}{2N-1}, \nonumber \\
 && F_{mn} = {\cal Q} \epsilon_{mn}.
 \end{eqnarray}
 The second term in (\ref{2eq32}) is always dominant as $r\to \infty$.
 The  behavior of the solution depends on the characteristic curvature
 $k$. The properties of the solution are summarized in Tab.~IV.

 When $k=1$, the causal structure is as follows. (i) When
 $0<N<1/2$, the cosmological constant is positive; the second term
 is negative. Therefore there will be a cosmological horizon regardless
  of the sign of the mass parameter ${\cal M}$. (ii) When
 $1/2<N<1 $, the cosmological constant has to be negative. In that
 case, black hole horizon appears when ${\cal M}>0$. The mass of
 black hole mass is given by (\ref{2eq17}). If ${\cal M}<0$, the
 singularity at $r=0$ becomes naked. In this case, the black hole
 (cosmological) horizon is a positive constant curvature
 hypersurface.

 When $k=-1$, the solution describes a black hole if $0<N<1/2$ and
 ${\cal M}>0$. On the other hand, in the case of $1/2<N<1$, the
 cosmological constant is positive and the second term is always
 negative. Therefore there is a cosmological horizon. Here the
 black hole (cosmological) horizon is a negative constant
 curvature hypersurface.

 This solution cannot be uplifted to $(D+d)$ dimensions according
 to (\ref{2eq2}) unless ${\cal Q}=0$ or $\Lambda_0=0$.  The case of ${\cal Q}=0$
 results in the higher dimensional solution (\ref{2eq24}) or
 (\ref{2eq25}) with $D=4$, while the higher dimensional solution
 is included in (\ref{2eq39}) as a special case when
 $\Lambda_0=0$.
\begin{table}[ht]
\caption{The property of the solution (\ref{2eq32}) for different
parameters. }
 \label{tab:4}
\begin{center}
\begin{tabular}{l|c|c|c|c}
\hline \hline
 $k$ & $1$& $1$ & $-1$& $-1$ \\
 \hline
  $N$ & $0<N<1/2$ & $1/2 <N< 1$& $ 0<N<1/2$ & $1/2 <N<1$ \\
 \hline
 $\Lambda_0 $ & $>0$  & $ <0 $ & $ <0 $ & $>0$ \\
  \hline
  Solution &  CH
   & BH (${\cal M}>0$) &
   BH (${\cal M} >0$) & CH   \\
\hline \hline
\end{tabular}
\end{center}
\end{table}

 \item Solution 2.
 \begin{eqnarray}
 \label{2eq34}
 && A =-\frac{\cal M}{r^{2N-1}} -\frac{\Lambda_0
      e^{a\phi_0}}{N(4N-1)}r^{2N} +\frac{k}{r^{2N-2}}, \\
&& \phi =\phi_0 -\sqrt{4N(1-N)}\ln r, \ \ \ R=r^N, \nonumber \\
&& N =\frac{1}{1+a^2}, \ \ \ b =\frac{Na}{1-N}=\frac{1}{a}, \nonumber \\
&& {\cal Q}^2 = 4 (1-N)ke^{b\phi_0}, \nonumber \\
\label{2eq35}
 && F_{mn} = {\cal Q} \epsilon_{mn}.
\end{eqnarray}
This solution is interesting. It follows from (\ref{2eq35}) that
 $k>0$ since ${\cal Q}^2 >0$. From the expression
(\ref{2eq34}), we can see that the third term in (\ref{2eq34}) is
dominant as $0<N<1/2$, while the domination term is the second
term if $1/2<N<1$ as $r\to \infty$.  (i) When $0<N<1/4$, the
solution always describes a black hole in spite of the sign of the
parameter ${\cal M}$ if $\Lambda_0<0$; if $\Lambda_0
>0$, one has to have ${\cal M}>0$ in order to have a black hole
horizon. In other cases the solution describes a naked
singularity. (ii) When $1/4 <N<1/2$, the solution also describes a
black hole in spite of the sign of ${\cal M}$ if $\Lambda_0 >0$.
However, if $\Lambda_0<0$, the mass parameter has to be positive,
${\cal M}>0$. (iii) When $1/2<N<1$, the solution has a
cosmological horizon if $\Lambda_0 >0$ in spite of the sign of
${\cal M}$; if $\Lambda_0 <0$, it is a black hole solution
provided ${\cal M}>0$. In other cases, the singularity at $r=0$ is
naked. The result is summarized in Tab.~V. The solution
(\ref{2eq34}) cannot be uplifted to a higher dimension according
to (\ref{2eq2}) unless ${\cal Q}=0$ or $\Lambda_0 =0$. For  the
former, the uplifting solution is just the one (\ref{2eq20}) with
$D=4$. On the other hand, if $\Lambda_0=0$, the resulting higher
dimensional solution is included in (\ref{2eq39}) below.
\begin{table}[ht]
\caption{The property of the solution (\ref{2eq34}) for different
parameters. }
 \label{tab:5}
\begin{center}
\begin{tabular}{l|c|c|c|c|c|c}
\hline \hline
  $N$ & $0<N<1/4$ & $0<N<1/4 $ & $1/4 <N< 1/2$ & $1/4<N<1/2$ & $ 1/2<N<1$ & $1/2 <N<1$ \\
 \hline
 $\Lambda_0 $ & $>0$  & $ <0 $ & $ >0 $ & $<0$ & $>0$& $<0$  \\
  \hline
  Solution &  BH (${\cal M}>0$) & BH & BH
   & BH (${\cal M}>0$) & CH &
   BH (${\cal M} >0$)   \\
\hline \hline
\end{tabular}
\end{center}
\end{table}

\item Solution 3.
\begin{eqnarray}
\label{2eq36}
 && A =-\frac{\cal M}{r^{2N-1}} -\frac{\Lambda_0
               e^{a\phi_0}}{1-N}r^{2-2N}
        +\frac{{\cal Q}^2 e^{-b\phi_0}}{4(1-N)} r^{2-2N}, \\
&& \phi =\phi_0 -\sqrt{4N(1-N)}\ln r, \ \ \ R=r^N, \nonumber \\
&& N = \frac{a^2}{1+a^2}, \ \ \ b=a, \nonumber \\
&& {\cal Q}^2 e^{-b\phi_0}= 4(1-N)k +4\Lambda_0 (2N-1)
e^{a\phi_0}, \nonumber \\
\label{2eq37}
 && F_{mn} = {\cal Q} \epsilon_{mn}
\end{eqnarray}
The solution (\ref{2eq36}) can be reexpressed as
\begin{equation}
\label{2eq38}
 A = -\frac{{\cal M}}{r^{2N-1}}+\frac{1}{2N-1}\left
(k -\frac{{\cal Q}^2 e^{-b\phi_0}}{2}\right) r^{2-2N}.
\end{equation}
The second term in (\ref{2eq38}) is always dominant as $r\to
\infty$. Therefore when the coefficient in front of $r^{2-2N}$ is
positive, the solution describes a black hole if ${\cal M}>0$,
otherwise a naked singularity. On the other hand, when the
coefficient in front of $r^{2-2N}$ is negative, the solution has a
cosmological horizon if ${\cal M}<0$, otherwise it is a naked
singularity solution. When $k=0$ this solution reduces to the
solution 2 for the case of $k=0$.

Note that this set of solutions has $b=a$, satisfying the relation
of (\ref{2eq28}). Therefore we can uplift this solution to the
case in $(D+d)$ dimensions according to (\ref{2eq2}). Redefining
$r^{1/(d+1)} \to r$ and rescaling coordinates and parameters
${\cal M}$ and ${\cal Q}$, the resulting solution can be expressed
as
\begin{eqnarray}
\label{2eq39}
 && ds^2_{4+d}= -f dt^2 +f^{-1} dr^2 +r^2
\delta_{ij}dy^idy^j +
     {\cal B}^2 h_{mn}dx^mdx^n, \nonumber \\
&& F_{mn}={\cal Q} \epsilon_{mn},
\end{eqnarray}
where
\begin{eqnarray}
\label{2eq40}
 && f= -\frac{\cal M}{r^{d-1}} -\frac{2 \Lambda_{\rm eff}
         }{d(d+1)}r^2, \nonumber \\
 && {\cal B}^2 = \frac{1}{4\Lambda_0}\left ((d+2)k
           \pm \sqrt{(d+2)^2k^2 -4\Lambda_0 (d+1) {\cal Q}^2}\right),
          \nonumber \\
 && \Lambda_{\rm eff} = \frac{d}{d+2}\Lambda_0
          -\frac{d}{4(d+2)}\frac{{\cal Q}^2}{{\cal B}^4},
 \end{eqnarray}
 and $h_{mn}dx^mdx^n$ denotes a two-dimensional space with
 constant curvature $2k$. The solution is valid only when ${\cal B}^2>0$.
 To our best knowledge, this solution is new.
 The solution (\ref{2eq39}) is nothing
 but a $(d+2)$-dimensional asymptotically AdS black hole (dS) solution
 with conformal Ricci flat black hole (cosmological) horizon, times a two
 dimensional space with constant curvature $2k$ with
 ${\cal M} >(<)0$. One can
 see from (\ref{2eq40}) that the effective cosmological constant
 depends on the original cosmological constant, magnetic charge
 ${\cal Q}$ and the curvature $k$. When ${\cal M}=0$ and
 $\Lambda_{\rm eff} <0$, this is a spacetime $AdS_{d+2} \times
 S^2$, known as a Freund-Rubin type solution~\cite{FR}.

\end{itemize}

\subsubsection{Electric charged solution}

For completeness, here we present and analyze the electric charged
solution to equations (\ref{2eq29}). For electric charged solution
we need not restrict to $D=4$. Of course, the solutions presented
below cannot be uplifted to $(D+d)$ dimensions according to
(\ref{2eq2}).

When $k=0$, we find two sets of solutions.
\begin{itemize}
\item Solution 1.
\begin{eqnarray}
\label{2eq41}
 && A = -\frac{\cal M}{r^{(D-2)N-1}}-
     \frac{2\Lambda_0 e^{a\phi_0}r^{2N}}{N(D-2)(D N-1)}
     + \frac{{\cal Q}^2 e^{b\phi_0}}{2N(D-2)((D-4)N+1)r
     ^{2(D-3)N}}, \\
&& \phi= \phi_0 -\sqrt{2N(1-N)(D-2)}\ln r, \ \ \ R= r^N,
   \nonumber \\
&& N = \frac{2}{2+(D-2)a^2}, \ \ \ b=a,
    \nonumber \\
\label{2eq42} && F_{tr} =\frac{{\cal Q}
e^{b\phi_0}}{r^{(D-4)N+2}},
\end{eqnarray}
where ${\cal Q}$ is an integration constant related to the
electric charge $Q$ of the solution via $Q = {\cal Q} V_h/4\pi$.
The metric function $A$ is dominant by the first term if $
0<N<1/D$, while by the second term if  $ 1/D <N<1$ as $r \to
\infty$. (1) When $0<N<1/D$, the case of $\Lambda_0 <0$ describes
a black hole spacetime if ${\cal M}<0$, and a naked singularity
for ${\cal M}>0$. The case of $\Lambda_0 >0$ has a cosmological
horizon if ${\cal M}>0$, and the singularity becomes naked if
${\cal M}<0$.  (ii) When $1/D <N<1$, if the cosmological constant
is positive, the solution has a cosmological horizon regardless of
the sign of ${\cal M}$. On the other hand, if the cosmological
constant is negative, the solution describes a black hole as
${\cal M}>0$, otherwise a naked singularity, once again. The
result is summarized in Tab.~VI. The cases of naked singularity
are not included there. In addition, we mention here that due to
the electric charge, it is possible to have two black hole
horizons.
\begin{table}[ht]
\caption{The property of the solution (\ref{2eq41}) for different
parameters. }
 \label{tab:6}
\begin{center}
\begin{tabular}{l|c|c|c|c}
\hline \hline
  $N$ & $0<N<1/D$ & $0<N<1/D $ & $1/D <N< 1$ & $1/D<N<1$  \\
 \hline
 $\Lambda_0 $ & $>0$  & $ <0 $ & $ >0 $ & $<0$  \\
  \hline
  Solution &  CH (${\cal M}>0$) & BH  (${\cal M}<0$) & CH
   & BH (${\cal M}>0$) \\
\hline \hline
\end{tabular}
\end{center}
\end{table}

\item Solution 2.
\begin{eqnarray}
\label{2eq43}
 && A = -\frac{\cal M}{r^{(D-2)N-1}} +\frac{{\cal Q}^2
            e^{b\phi_0}r^{2-2N}}{2 (1-2N)((D-4)N+1)},
            \\
 && \phi=\phi_0 -\sqrt{2N(1-N)(D-2)}\ln r, \ \ \ R= r^N,
   \nonumber \\
 && N = \frac{(D-2)a^2}{2+(D-2)a^2}, \ \ \ b=-(D-3)a, \nonumber \\
 && 2 \Lambda_0 e^{a\phi_0}= \frac{N(D-4)+1}{2(2N-1)}
      {\cal Q}^2 e^{b\phi_0}, \nonumber \\
 && F_{tr}= \frac{{\cal Q} e^{b\phi_0}}{r^{N(4-D)}}.
 \end{eqnarray}
 The metric function $A$ is always dominant by the second term as
 $r\to \infty$. Therefore when $0<N<1/2$, the cosmological
 constant must be negative, and the solution has a black hole
 horizon if ${\cal M}>0$, otherwise it is a naked singularity.
 On the other hand, when $1/2 <N<1$, the cosmological constant
 should be positive. In this case, the solution has a cosmological
 horizon if ${\cal M}<0$, otherwise, the singularity at $r=0$
 becomes naked, again. The property of the solution is summarized
 in Tab.~VII.
\begin{table}[ht]
\caption{The property of the solution (\ref{2eq43}) for different
parameters. }
 \label{tab:7}
\begin{center}
\begin{tabular}{l|c|c|c|c}
\hline \hline
  $N$ & $0<N<1/2$ & $0<N<1/2$ & $1/2 <N< 1$ & $1/2<N<1$  \\
 \hline
 $\Lambda_0 $ & $<0$  & $ <0 $ & $ >0 $ & $>0$  \\
  \hline
  ${\cal M}$ & $>0$ &$<0$& $>0$ & $<0$ \\
  \hline
  Solution &  BH  & NK &  NK  & CH \\
\hline \hline
\end{tabular}
\end{center}
\end{table}

\end{itemize}

When $k\ne 0$, we find three sets of solutions. They are
\begin{itemize}
\item Solution 1.
\begin{eqnarray}
\label{2eq45}
&& A = -\frac{\cal M}{r^{(D-2)N-1}}
           +\frac{(D-3)k r^{2-2N}}{(2N-1)((D-4)N+1)}
           +\frac{{\cal Q}^2 e^{b\phi_0}r^{-2(D-3)N}}{2N(D-2)
           ((D-4)N+1)}, \\
&& \phi =\phi_0 -\sqrt{2N(1-N)(D-2)}\ln r, \ \ \ R=r^N,
   \nonumber \\
&& N= \frac{(D-2)a^2}{2+(D-2)a^2}, \ \ \ b =\frac{2}{(D-2)a},
     \nonumber \\
&& 2\Lambda_0 e^{a\phi_0}= -\frac{(D-2)(D-3)(1-N)k}{(2N-1)},
      \nonumber \\
&& F_{tr}=\frac{{\cal Q}e^{b\phi_0}}{r^{(D-4)N+2}}.
\end{eqnarray}
For this solution, one can see that the second term in
(\ref{2eq45}) always dominates over other two terms as $r\to
\infty$. (i) When $k=1$, the cosmological constant has to be
positive for $ 0<N<1/2$, the solution  always has a cosmological
horizon regardless of the sign of ${\cal M}$; for the case $1/2
<N<1$, the cosmological constant is negative. In that case  black
hole horizon appears if ${\cal M}>0$, otherwise it is a naked
singularity solution. (ii) When $k=-1$, the solution always has  a
cosmological horizon if $1/2 <N<1$, while it describes a black
hole in the case of $0<N<1/2$ with ${\cal M}>0$. The result is
summarized in Tab.~VIII.
\begin{table}[ht]
\caption{The property of the solution (\ref{2eq45}) for different
parameters. }
 \label{tab:8}
\begin{center}
\begin{tabular}{l|c|c|c|c}
\hline \hline
  $k$ & $1$ & $1$ & $-1$ & $-1$ \\
  \hline
 $N$ & $0<N<1/2$ & $1/2<N<1$ & $0 <N< 1/2$ & $1/2<N<1$  \\
 \hline
 $\Lambda_0 $ & $>0$  & $ <0 $ & $ <0 $ & $>0$  \\
  \hline
  Solution &  CH  & BH (${\cal M}> 0$) &  BH (${\cal M}>0$) & CH \\
\hline \hline
\end{tabular}
\end{center}
\end{table}

\item Solution 2.
\begin{eqnarray}
\label{2eq47}
 && A =-\frac{\cal M}{r^{(D-2)N-1}}
       -\frac{2\Lambda_0 e^{a\phi_0}r^{2N}}{N (D-2)(ND-1)}
       +\frac{(D-3){\cal Q}^2 e^{b\phi_0}}
       {2(1-N)(D-2)((D-4)N+1)r^{2N-2}},
       \\
&& \phi =\phi_0 -\sqrt{2N(1-N)(D-2)}\ln r, \ \ \ R=r^N,
   \nonumber \\
&& N = \frac{2}{2+(D-2)a^2}, \ \ \ b= -\frac{2(D-3)}{(D-2)a},
\nonumber \\
&& {\cal Q}^2 e^{b\phi_0}= \frac{2(D-3)(D-2)(1-N)k}{(D-4)N+1},
\nonumber \\
&& F_{tr}= \frac{{\cal Q }e^{b\phi_0}}{r^{N(4-D)}}.
\end{eqnarray}
For nonvanishing electric charge, the solution holds only for the
case of $k=1$. When $0<N<1/2$, the third term in (\ref{2eq47}) is
always dominant over other terms as $r\to \infty$. But the second
term will changes its sign around $N=1/D$. So we have three cases.
(i) When $0<N<1/D$, a black hole horizon always appears if
$\Lambda_0 <0$, otherwise in order to have a black hole horizon,
the mass parameter must be positive. (ii) When $1/D <N<1/2$, a
black hole horizon always appears if $\Lambda_0
>0$. Otherwise the mass parameter has to be positive. (iii) When
$1/2 <N<1$, the second term in (\ref{2eq47}) becomes dominant. In
this case, when $\Lambda_0 >0$, the solution always has a
cosmological horizon; when $\Lambda_0 <0$ a black hole horizon
appears  provided ${\cal M}>0$. The property of the solution
(\ref{2eq47}) is summarized in Tab.~IX.
\begin{table}[ht]
\caption{The property of the solution (\ref{2eq47}) for different
parameters. }
 \label{tab:9}
\begin{center}
\begin{tabular}{l|c|c|c|c|c|c}
\hline \hline
  $N$ & $0<N<1/D$ & $0<N<1/D $ & $1/D <N< 1/2$ & $1/D<N<1/2$ & $ 1/2<N<1$ & $1/2 <N<1$ \\
 \hline
 $\Lambda_0 $ & $>0$  & $ <0 $ & $ >0 $ & $<0$ & $>0$& $<0$  \\
  \hline
  Solution &  BH (${\cal M}>0$) & BH & BH
   & BH (${\cal M}>0$) & CH &
   BH (${\cal M} >0$)   \\
\hline \hline
\end{tabular}
\end{center}
\end{table}

\item Solution 3.
\begin{eqnarray}
\label{2eq49}
 && A = -\frac{\cal M}{r^{(D-2)N-1}}
         -\frac{( 2\Lambda_0 e^{a\phi_0} -(D-3){\cal Q}^2
  e^{b\phi_0}/2)r^{2-2N}}{(1-N)(D-2)((D-4)N+1)}, \\
&&  \phi =\phi_0 -\sqrt{2N(1-N)(D-2)}\ln r, \ \ \ R=r^N,
  \nonumber \\
&&  N= \frac{a^2}{2 +(D-2)a^2}, \ \ \ b=-(D-3)a, \nonumber \\
&& 2\Lambda_0 e^{a\phi_0} =-\frac{(D-2)(1-N)}{2N-1}\left ((D-3)k -
    \frac{(D-4)N+1}{2(D-2)(1-N)}{\cal Q}^2 e^{b\phi_0}\right),
    \nonumber \\
&& F_{tr}= \frac{{\cal Q }e^{b\phi_0}}{r^{N(4-D)}}.
\end{eqnarray}
The metric function $A$ is always dominant by the second term in
(\ref{2eq49}) as $r \to \infty$. When ${\cal M}>0$, the solution
has a black hole horizon if $\triangle \equiv 2\Lambda_0
e^{a\phi_0} -(D-3){\cal Q}^2 e^{b\phi_0}/2 <0$, otherwise, it is a
naked singularity solution. On the other hand, when ${\cal M}<0$
the solution has a cosmological horizon if $\triangle >0$, while
the singularity becomes naked again if $\triangle <0$.

\end{itemize}


\section{Dimensional reduction: curved case}
 In the dimensional reduction from (\ref{2eq1}) to (\ref{2eq3}),
 we assumed $ds_d^2$ is a $d$-dimensional Ricci flat Euclidean
 space. Now we relax this condition and suppose it is a nonzero
 constant curvature space with curvature scalar
 \begin{equation}
 \label{3eq1}
 R_d = d(d-1) k_d.
 \end{equation}
 Without loss of generality, one may take $k_d= \pm 1$. If
 $k_d=0$, the reduction goes back to the case discussed in the
 previous section. Doing the reduction along $y^i$ in (\ref{2eq2}),
 we obtain
 \begin{equation}
 \label{3eq2}
 S=\frac{V_q}{16\pi G}\int d^Dx\sqrt{-g}\left(R
 -\frac{1}{2}(\partial \phi)^2 -2\Lambda_0 e^{a\phi}
 -2\Lambda_1e^{c\phi}\right),
 \end{equation}
 where the relation between $\phi$ and $\alpha$ is the same as
 (\ref{2eq4}),  $a$ is still given by (\ref{2eq6}) and
 \begin{equation}
 \label{3eq3}
 c=-\frac{D+d-2}{d}a, \ \ \ 2 \Lambda_1 = -d(d-1)k_d.
 \end{equation}
 The action (\ref{3eq2}) describes an Einstein-dilaton gravity theory
 with two Liouville-type dilaton potentials provided $a \ne c$.
 Otherwise, the two terms can be reduced to one. Note that there is a symmetry,
  \begin{equation}
  \label{3eq4}
  a \leftrightarrow c\ \ \ \ {\rm and}\ \ \ \Lambda_0 \leftrightarrow \Lambda_1,
  \end{equation}
  in the action (\ref{3eq2}). Varying the action
 (\ref{3eq2}), we have the equations of motion
 \begin{eqnarray}
 \label{3eq5}
 && R_{\mu\nu}=\frac{1}{2}\partial_{\mu}\phi \partial_{\nu}\phi
    +\frac{2}{D-2}\Lambda_0 e^{a\phi}g_{\mu\nu}+\frac{2}{D-2}\Lambda_1
    e^{c\phi}g_{\mu\nu}, \nonumber \\
  && \nabla^2\phi -2a\Lambda_0 e^{a\phi} -2c\Lambda_1e^{c\phi}=0.
  \end{eqnarray}
  As the above, we now solve the equations of motion with the ansatz (\ref{2eq8}),
  and assuming that $\Lambda_0$, $\Lambda_1$, $a$ and $c$ are arbitrary constants.
  Once again, the solutions depend on the curvature $k$.

  (1) When $k=0$, we find that the equations (\ref{3eq5}) have a
  consistent solution only as $a=c$. The solution is
  \begin{eqnarray}
  \label{3eq6}
&& A(r)= -\frac{{\cal M}}{r^{(D-2)N-1}} -\frac{2 (\Lambda_0
+\Lambda_1)
 e^{a\phi_0}r^{2N}}{N(ND-1)(D-2)}, \nonumber
   \\
 && \phi =\phi_0 -\sqrt{2N(1-N)(D-2)}\ln r, \nonumber \\
 && R =r^N, \ \ \ c=a, \nonumber \\
 && N =\frac{2}{2+c^2(D-2)}.
 \end{eqnarray}
 Therefore the property of this solution is the same as the one
 (\ref{2eq15}). Only difference is that $\Lambda_0$ in
 (\ref{2eq15}) is replaced by $\Lambda_0 +\Lambda_1$ here. Because
 the relation between $a$ and $c$ in (\ref{3eq6}) does not satisfy
 (\ref{3eq3}), the solution therefore cannot be
 uplifted to $(D+d)$ dimensions unless $\Lambda_1=0$ or $\Lambda_0=0$.
 When $\Lambda_1=0$, the resulting solution is just the one in
 (\ref{2eq20}). Here we discuss the case of $\Lambda_0=0$. Note
 that in this case, one does not need the condition $c=a$ in
 (\ref{3eq6}). Uplifting the solution, we find that the metric can
 be expressed as, upon some rescalings of the coordinates and the
 parameters ${\cal M}$,
 \begin{equation}
 \label{3eq7}
 ds^2_{D+d}= -\left (k_d -\frac{{\cal M}}{r^{d-1}}\right)dt^2
        +\left (k_d -\frac{{\cal M}}{r^{d-1}}\right)^{-1} dr^2 +r^2
        h_{ij} dy^idy^j + \delta_{mn}dx^mdx^n.
 \end{equation}
 This is nothing but a $(d+2)$-dimensional Schwarzschild metric
 times a $(D-2)$-dimensional Ricci flat space. Here $h_{ij}dy^idy^j$
 denotes the line element of a $d$-dimensional
 space with curvature scalar $d(d-1)k_d$.

 (2) When $k\ne 0$, we find three sets of solutions to the
 equations (\ref{3eq5}) of motion.
\begin{itemize}
 \item Solution 1.
\begin{eqnarray}
\label{3eq8}
  && A =-\frac{{\cal M}}{r^{(D-2)N -1}} +\frac{(D-3)k
         r^{2-2N}}{(2N-1) (N(D-4) +1)}-\frac{2\Lambda_1
         e^{c\phi_0}r^{2N}} {N (D-2)(ND-1)}, \\
&& \phi= \phi_0 -\sqrt{2N(1-N)(D-2)}\ln r, \ \ \  R= r^N, \nonumber \\
&& N= \frac{a^2(D-2)}{2+a^2 (D-2)}, \ \ \  c =\frac{2}{(D-2)a},
        \nonumber \\
\label{3eq9}
 && 2\Lambda_0 =-\frac{(D-3)(D-2) (1-N)k
e^{-a\phi_0}}{2N-1}.
\end{eqnarray}
Let us first consider the case of $k=1$. As $r\to \infty$, the
second term in (\ref{3eq8}) is dominant if $ 0<N<1/2$, while the
third term does if $1/2 <N<1$.
 (i) In the range $0<N<1/D$, the cosmological constant $\Lambda_0$
 must be positive. In this case, the solution must have a
 cosmological horizon if $\Lambda_1 >0$, regardless of the sign of
 the mass parameter; on the other hand, in order to have a
 cosmological horizon, the mass parameter must be negative if
 $\Lambda_1 <0$. (ii) In the range $1/D <N<1/2$, if $\Lambda_1<0$, the solution has
 a cosmological horizon  regardless of  the sign of ${\cal M}$; to have a cosmological
 horizon, ${\cal M}$ must be negative if $\Lambda_1>0$. (iii) If $1/2 <N<1$, the
 cosmological constant $\Lambda_0$ is negative. When $\Lambda_1 >0$
 the solution has a
 cosmological horizon  regardless of the sign of ${\cal M}$, while black hole
 horizon appears provided
 ${\cal M}>0$ when $\Lambda_1 <0$. In other cases, the singularity
 at $r=0$ is naked. The result is summarized in Tab.~X. This
 solution cannot be uplifted to $(D+d)$ dimensions according to
 (\ref{2eq2}) unless $\Lambda_0=0$ or $\Lambda_1=0$. For the
 latter, the resulting solution is (\ref{2eq24}). For the case
 of $\Lambda_0=0$,  the higher dimensional solution turns out to
 be (\ref{3eq7}).

\begin{table}[ht]
\caption{The property of the solution 1 in (\ref{3eq8}) for
different parameters in the case of $k=1$. }
 \label{tab:10}
\begin{center}
\begin{tabular}{l|c|c|c|c|c|c}
\hline \hline
 $N$ & $0<N<1/D$ & $ 0<N<1/D$ & $1/D <N< 1/2$  & $1/D <N<1/2$
      & $ 1/2 <N<1$ & $ 1/2<N<1 $ \\
 \hline
 $\Lambda_0 $ & $>0$ &$>0$ & $>0$  & $ >0 $ & $ <0 $ &$<0$ \\
  \hline
 $\Lambda_1$ & $>0$ &$<0$& $>0$ &$<0$ &$ >0$ &$<0$ \\
 \hline
  Solution &  CH & CH (${\cal M}<0$) &
   CH & CH(${\cal M}<0$) &  CH & BH (${\cal M}>0$)  \\
\hline \hline
\end{tabular}
\end{center}
\end{table}

When $k=-1$, the cosmological constant $\Lambda_0$ is negative in
the range $0<N<1/2$, while it is positive in the range $1/2 <N<1$.
(i) In the range $0<N<1/D$, a black hole horizon always appears if
$\Lambda_1 <0$  regardless of the sign of ${\cal M}$. If
$\Lambda_1>0$, one has to have ${\cal M}>0$ in order to have a
black hole horizon. (ii) In the range of $1/D <N<1/2$, the
solution always has a black hole horizon if $\Lambda_1>0$,
otherwise the mass parameter must be positive if $\Lambda_1<0$ in
order to have a black hole horizon. (iii) In the range of $1/2
<N<1$, a cosmological horizon will appear if $\Lambda_1>0$ and
${\cal M}<0$; if $ \Lambda_1<0$, the solution always describes a
black hole spacetime. The property of the solution (\ref{3eq8}) is
summarized in Tab.~XI in the case of $k=-1$. Once again, because
the relation between $a$ and $c$ in the solution (\ref{3eq8}) does
not satisfy the one in (\ref{3eq3}), one cannot uplift the
solution to higher dimensional case according to (\ref{2eq2})
unless $\Lambda_0=0$ or $\Lambda_1=0$. In the latter case, the
uplifting solution is (\ref{2eq25}); while the solution turns out
to be (\ref{3eq7}), once again, for the case of $\Lambda_0=0$.

\begin{table}[ht]
\caption{The property of the solution 1 in (\ref{3eq8}) for
different parameters in the case of $k=-1$. }
 \label{tab:11}
\begin{center}
\begin{tabular}{l|c|c|c|c|c|c}
\hline \hline
 $N$ & $0<N<1/D$ & $0<N<1/D$ & $1/D <N< 1/2$
    & $1/D <N< 1/2$& $ 1/2 <N<1$ & $1/2 <N<1$ \\
 \hline
 $\Lambda_0 $ & $<0$  & $ <0 $ & $<0$ &$<0$ &$>0$ & $ >0 $ \\
  \hline
 $\Lambda_1$ & $>0$ &$<0$ & $>0$ &$<0$ & $>0$ &$ <0 $ \\
 \hline
  Solution &  BH(${\cal M}>0$)
   & BH  & BH  & BH (${\cal M}>0$) & CH (${\cal M}<0$)& BH \\
\hline \hline
\end{tabular}
\end{center}
\end{table}

\item Solution 2.
\begin{eqnarray}
\label{3eq10}
  && A =-\frac{{\cal M}}{r^{(D-2)N -1}} +\frac{(D-3)k
         r^{2-2N}}{(2N-1) (N(D-4) +1)}-\frac{2\Lambda_0
         e^{a \phi_0}r^{2N}} {N (D-2)(ND-1)}, \\
&& \phi= \phi_0 -\sqrt{2N(1-N)(D-2)}\ln r, \ \ \
  R= r^N, \nonumber \\
&& N= \frac{c^2(D-2)}{2+c^2 (D-2)}, \ \ \  a =\frac{2}{(D-2)c},
        \nonumber \\
\label{3eq}
 && 2\Lambda_1 =-\frac{(D-3)(D-2) (1-N)k
e^{-c\phi_0}}{2N-1}.
\end{eqnarray}
This is symmetric to solution 1 according to (\ref{3eq4}). We do
not therefore present the analysis of the solution here.

\item Solution 3.
\begin{eqnarray}
\label{3eq12}
 && A(r)= -\frac{{\cal M}}{r^{(D-2)N-1}} + \frac{(D-3)k
 r^{2-2N}}{(2N-1)(N(D-4)+1)}, \\
 && \phi =\phi_0 -\sqrt{2N(1-N)(D-2)}\ln r, \ \ \
  R=r^N, \nonumber \\
 && N= \frac{c^2 (D-2)}{2+c^2(D-2)}, \ \ \ c=a, \nonumber\\
 \label{2eq}
 && 2\Lambda_0+2\Lambda_1 = -\frac{(D-3)(D-2) (1-N)k e^{-c\phi_0}}{2N-1}.
 \end{eqnarray}
The solution is the same as the one (\ref{2eq21}) for the case
with one Liouville-type potential. The only difference is
$\Lambda_0$ there is replaced by $\Lambda_0+\Lambda_1$ here. The
causal structure of the solution is of course the same as the one
for the solution (\ref{2eq21}). In addition, due to $c=a$, we also
cannot uplift the solution to $D+d$ dimensions according to
(\ref{2eq2}) unless $\Lambda_1=0$ or $\Lambda_0 =0$. For the
former, resulting solution is given by (\ref{2eq24}) or
(\ref{2eq25}) depending on the sign of  $\Lambda_0$. For the
latter, however, we find that the $(D+d)$-dimensional solution can
be expressed as
\begin{equation}
\label{3eq14}
 ds^2_{D+d} = -\left(k -\frac{\cal
M}{r^{D+d-3}}\right) dt^2
      +\left(k -\frac{\cal M}{r^{D+d-3}}\right)^{-1}
      + r^2( q_{ij}dy^idy^j +h_{mn}dx^mdx^n)),
\end{equation}
The same as the above, some rescalings of coordinates and the
parameter ${\cal M}$ have been made here, and the subspace
described by ($q_{ij}dy^idy^j +h_{mn}dx^mdx^n)$ has been
normalized to have constant curvature $(D+d-2)(D+d-3)k$. Note that
the uplifting holds only for the case where $k_b$ and $k$ have
same signs since in this case we have $N=(D+d-2)/(D+2d-2)>1/2$. It
can be immediately seen that the solution satisfies the vacuum
Einstein equations in $(D+d)$ dimensions and is asymptotically
flat.
\end{itemize}

When the Maxwell field is present, the resulting action will have
an additional term
\begin{equation}
\label{3eq15}
 S=\frac{V_q}{16\pi G} \int d^Dx\sqrt{-g}\left( R
-\frac{1}{2} (\partial \phi)^2 -2\Lambda_0 e^{a\phi} -2\Lambda_1
e^{c\phi}
 -\frac{1}{4}e^{-b\phi} F^2_2\right),
 \end{equation}
 where $a$ and $b$ are given by (\ref{2eq6}) and (\ref{2eq26}),
 respectively, and $c$ and $\Lambda_1$ are given by (\ref{3eq3}).
 Varying the action, we have the equations of motion
 \begin{eqnarray}
 \label{3eq16}
&& R_{\mu\nu}=\frac{1}{2}\partial_{\mu}\phi \partial_{\nu}\phi
    +\frac{2}{D-2}\Lambda_0 e^{a\phi}g_{\mu\nu}+\frac{2}{D-2}\Lambda_1
    e^{c\phi}g_{\mu\nu} +\frac{1}{2}e^{-b\phi}\left(F_{\mu\lambda}F_{\nu}^{\ \lambda}
    -\frac{1}{2(D-2)}F_2^2g_{\mu\nu}\right), \nonumber \\
&& \partial_{\mu}(\sqrt{-g} e^{-b\phi} F^{\mu\nu})=0, \ \ \
   F_{\mu\nu,\lambda}+F_{\nu\lambda,\mu} +F_{\lambda\mu,\nu}=0,
  \nonumber \\
  && \nabla^2\phi -2a\Lambda_0 e^{a\phi} -2c\Lambda_1e^{c\phi}+\frac{b}{4}e^{-b\phi}F_2^2=0.
  \end{eqnarray}
 As the case of one Liouville-type potential,
 it is straightforward to find and
 analyze the magnetic/electric charged solutions to equations (\ref{3eq16})
 of motion. But the expressions of solutions and analysis are
  rather complicated and we shall not present them here.

\section{Conclusions and Discussion}

By dimensionally reducing an Einstein-Maxwell theory with a
cosmological constant to a lower dimension, we can obtain an
Einstein-Maxwell-dilaton theory with one or two dilaton potentials
of Liouville-type. We have looked for and analyzed in some details
exact (neutral/magnetic/electric charged) solutions of these
theories. These found solutions have a rich structure depending on
the parameters in the theory. These solutions could have black
hole horizon(s), cosmological horizon, or a naked singularity.
These horizons can be a positive, zero or negative constant
curvature hypersurface. In particular, we noted that some black
hole solutions have negative gravitational mass. In addition these
solutions are neither asymptotically flat nor asymptotically
AdS/dS. Our study generalized existing results concerning
non-asymptotically AdS/dS solutions in the literature in some
directions. For example, (1) in our discussions, spacetime
dimensions $D$ and $d$ are almost arbitrary; (2) black hole
horizon(s) or cosmological horizon can be a hypersurface with
positive, zero or negative constant curvature; (3) the scalar
potential could be one or two Liouville-type term(s). Here it
would be worth mentioning some of existing literatures about
non-asymptotically AdS/dS solutions: the authors of \cite{Mann}
looked for solutions of Einstein-Maxwell-dilaton theory with
dilaton potential, but it is restricted to the case $k=1$; Chan in
\cite{Chan} discussed the modification of three dimensional BTZ
black hole by a scalar potential; the authors of
\cite{Caizhang,CJS} studied four dimensional topological dilaton
black hole solutions with $k=0$ or $-1$. Klemm in \cite{Klemm}
found supersymmetric solution in the four dimensional gauged
$N=4$, $SU(2)\times SU(2)$ supergarvity; Refs.~\cite{CR,CO,CZ,CMZ}
investigated domain wall solutions with $k=0$ in Einstein-dilaton
theory with a Liouville-type potential; Cvetic et al in
\cite{Cvetic} found a domain wall solution with $k=0$ in an
Einstein-Maxwell-dilaton theory with special coupling parameters
$a=b=\sqrt{2/p}$ in (\ref{2eq27}); for more recent discussions on
the non-asymptotically AdS/dS solutions with $k=1$ in four
dimensional Einstein-Maxwell-dilaton theory without dilaton
potential see \cite{CL,CGL}; other related discussions see
\cite{others}.

Since the action that we worked with could come from the
dimensional reduction from a higher dimensional Einstein-Maxwell
theory with a cosmological constant,  some of the found solutions
can be uplifted to the case of higher dimensions. We found that
some of these neither asymptotically flat nor AdS/dS solutions
have a higher dimensional origin; and their higher dimensional
origins have well-behaved asymptotics: they are either
asymptotically AdS/dS [see, for example, (\ref{2eq20}),
(\ref{2eq24}), (\ref{2eq25}) and (\ref{2eq39})], or asymptotically
flat (see, for example, (\ref{3eq7}) and (\ref{3eq14})]
with/without a compact constant curvature space. From the point of
view of holography, this observation is useful to better
understand the holographic properties of these non-asymptotically
AdS/dS solutions.

The black hole horizon and cosmological horizon of these solutions
have usual thermodynamical relations. For example, the Hawking
temperature of horizon is given by
\begin{equation}
T = \frac{1}{4\pi}|A'(r)|_{r=r_+,r_c},
\end{equation}
where $r_+$ and $r_c$ denote black hole horizon and cosmological
horizon, respectively. The horizon entropy is still by the
so-called area formula
\begin{equation}
S =\frac{V_qV_h}{4G} {\cal A},
\end{equation}
where ${\cal A}= r^{(D-2)N}|_{r=r_+,r_c}$, since we have worked in
the Einstein frame, where the area formula of black hole entropy
holds. With the black hole mass given by (\ref{2eq17}) or the
gravitational mass (\ref{2eq18}) for the cosmological horizon, it
is easy to show that the first law of thermodynamics holds for
both black hole horizon and cosmological horizon
\begin{equation}
\label{4eq3}
 {\rm d}M=T{\rm d}S +\Phi {\rm d}Q,
\end{equation}
where $Q$ is the electric charge or magnetic charge of the
solutions, and $\Phi$ is the corresponding chemical potential. As
an example, let us consider the electric charged solution given by
(\ref{2eq41}) and (\ref{2eq42}). In terms of black hole horizon
radius $r_+$ determined by the equation $A(r)|_{r=r_+}=0$, the
black hole mass is given by
\begin{equation}
M = \frac{ V_qV_h r_+^{(D-2)N-1}}{16\pi G } \left (
   -\frac{2\Lambda_0 e^{a\phi_0}r_+^{2N}}{ND-1}
   +\frac{{\cal Q}^2e^{b\phi_0}}{2((D-4)N+1)r_+^{2(D-3)N}}
   \right).
\end{equation}
The Hawking temperature $T$ and the entropy are
\begin{eqnarray}
&& T = \frac{1}{4\pi N(D-2) r_+} \left(-2 \Lambda_0
e^{a\phi_0}r_+^{2N} -\frac{{\cal Q}^2 e^{b\phi_0}}{2
r_+^{2(D-3)N}}
   \right); \nonumber \\
&& S = \frac{V_qV_h}{4G} r_+^{(D-2)N},
\end{eqnarray}
respectively. Note that in order to have a black hole horizon, the
constant $\Lambda_0$ has to be negative here so that the
temperature is always positive definiteness.  When $T=0$, it
indicates that the black hole is an extremal black hole with
vanishing Hawking temperature. The electric charge associated with
the black hole is
\begin{equation}
Q = \frac{V_h}{4\pi }{\cal Q}.
\end{equation}
And the conjugate chemical potential $\Phi$
\begin{equation}
\Phi = \frac{V_q {\cal Q} e^{b\phi_0}}{4G ((D-4)N+1)
r_+^{(D-4)N+1}}.
\end{equation}
These thermodynamic quantities satisfy the first law of black hole
thermodynamics (\ref{4eq3}). Indeed, as was shown by
Wald~\cite{Wald}, the first law of black hole mechanics always
holds for any stationary black hole spacetime in any gravitational
theory.

Finally we point out that the solutions we found are not
well-defined for some special values of $N$, for example, $N=1/D$
for the solution (\ref{2eq15}), (\ref{2eq41}), (\ref{2eq47}), and
(\ref{3eq6});  $N=1/2$ for (\ref{2eq21}), (\ref{in1}),
(\ref{2eq38}), (\ref{2eq43}), (\ref{2eq45}),(\ref{2eq49}),
(\ref{3eq8}), (\ref{3eq10}) and (\ref{3eq12}); and  $N=1/4$ for
(\ref{2eq30}) and (\ref{2eq34}). In fact for these special values,
we can also find solutions, but a term concerning the logarithmic
function of $r$ will appear in the metric function $A$. Here we do
not present those solutions, although the concerned discussions
are straightforward.

\section*{Acknowledgments}
We thank G. Clement for a carefully reading. One of the authors
(RGC) would like very much to express his gratitude to the Physics
Department, Baylor University for its hospitality. This work was
supported by Baylor University, a grant from Chinese Academy of
Sciences, a grant from NSFC, China (No. 10325525), and a grant
from the Ministry of Science and Technology of China (No.
TG1999075401).

\end{document}